\def\year{2017}\relax
\documentclass[letterpaper]{article}
\usepackage{aaai17}
\usepackage{times}
\usepackage{helvet}
\usepackage{courier}
\usepackage{graphicx}
\usepackage{multirow}
\usepackage{epstopdf}
\usepackage{comment}
\usepackage{caption}
\usepackage{subcaption}
\usepackage{CJKutf8} 
\usepackage[ruled]{algorithm2e}
\usepackage{booktabs}
\usepackage{amsmath}
\begin{document}
%
\title{Collaborative Company Profiling: Insights from an Employee's Perspective}

\author{Hao Lin$^{1}$, Hengshu Zhu$^{2}$\thanks{Corresponding Author.}, Yuan Zuo$^{1}$, Chen Zhu$^{2}$, Junjie Wu$^{1}$, Hui Xiong$^{3*}$\\
$^{1}$Beihang University, \{linhao2014, zuoyuan, wujj\}@buaa.edu.cn\\
$^{2}$Baidu Talent Intelligence Center,  \{zhuhengshu, zhuchen02\}@baidu.com\\
$^{3}$Rutgers University, hxiong@rutgers.edu\\
\\
}
\maketitle
\begin{abstract}
Company profiling is an analytical process to build an in-depth understanding of company's fundamental characteristics. It serves as an effective way to gain vital information of the target company and acquire business intelligence. Traditional approaches for company profiling rely heavily on the availability of rich finance information about the company, such as finance reports and SEC filings, which may not be readily available for many private companies. However, the rapid prevalence of online employment services enables a new paradigm --- to obtain the variety of company's information from their employees' online ratings and comments. This, in turn, raises the challenge to develop company profiles from an employee's perspective. To this end, in this paper, we propose a method named Company Profiling based Collaborative Topic Regression (CPCTR), for learning the latent structural patterns of companies. By formulating a joint optimization framework, CPCTR has the ability in collaboratively modeling both textual (e.g., reviews) and numerical information (e.g., salaries and ratings). Indeed, with the identified patterns, including the positive/negative opinions and the latent variable that influences salary, we can effectively carry out opinion analysis and salary prediction. Extensive experiments were conducted on a real-world data set to validate the effectiveness of CPCTR. The results show that our method provides a comprehensive understanding of company characteristics and delivers a more effective prediction of salaries than other baselines.

\end{abstract}

\section{Introduction}\label{sec:intro}
Recent years have witnessed the rapid development of technologies related to enterprise management, which can help organizations to keep up with the continuously changing business world. Along this line, a crucial demand is to build effective strategies for company profiling, which is an analytical process that results in an in-depth understanding of company's fundamental characteristics, and can therefore serve as an effective way to gain vital information of the target company and acquire business intelligence. With the help of profiling, a wide range of applications could be enabled including organization risk management~\cite{martin2007profiling}, enterprise integration~\cite{hollocks1997assessing}, and company benchmarking~\cite{knuf2000benchmarking,alling2002method,seong2008information,kerschbaum2008building,Zhu-KDD-2016}.

In the past decades, traditional approaches for company profiling rely heavily on the availability of the rich finance information about the company, such as finance reports and SEC filings, which may not be readily available for many private companies. Recently, with the rapid prevalence of online employment services, such as Glassgdoor, Indeed, and Kanzhun, a new paradigm is enabled for obtaining the variety of company's information from their (former) employees anonymously via the reviews, ratings and salaries of specific job positions. This, in turn, raises the question whether it is possible to develop company profiles from an employee's perspective. For example, we can help companies to identify their advantages and disadvantages, and to predict the expected salaries of different job positions for rival companies.

However, the heterogeneous characteristic of this public information imposes significant challenges to discover typical patterns of companies during profiling. To this end, in this paper we propose a model named Company Profiling based Collaborative Topic Regression (CPCTR) to formulate a joint optimization framework for learning the latent patterns of companies, which can collaboratively model both the textual information (e.g., review) and numerical information (e.g., salary and rating). With the identified patterns, including the positive/negative opinions and the latent variable that influences salary, we can effectively carry out opinion analysis and salary predictions for different companies. Finally, we conduct extensive experiments on a real-world data set. The results show that our algorithm provides a comprehensive interpretation of company characteristics and a more effective salary prediction than other baselines. Particularly, by analyzing the results obtained by CPCTR, many meaningful patterns and interesting discoveries can be observed, such as \emph{welfare and technology are the typical pros of Baidu, while those of Tencent are training and learning.}

\section{Related Work}\label{sec:related}
The related work of this paper can be grouped into two categories, namely \emph{topic modeling for opinion analysis} and \emph{matrix factorization for prediction}.

Probabilistic topic models are capable of grouping semantic coherent words into human interpretable topics. Archetypal topic models include probabilistic Latent Semantic Indexing (pLSI)~\cite{hofmann1999probabilistic} and Latent Dirichlet Allocation (LDA)~\cite{blei2003latent}. A lot of extensions have been proposed based on above standard topic models, such as author-topic model~\cite{rosen2004author}, correlated topic model (CTM)~\cite{lafferty2005correlated}, and dynamic topic model (DTM)~\cite{blei2006dynamic}, etc. Among them, numerous works focus on opinion analysis, especially for tackling the aspect-based opinion mining task~\cite{K2014Aspect,Zhu-ICDM-2014}. Moreover, a few works have attempted to combine ratings and review texts when performing opinion analysis~\cite{Ganu2009Beyond,Titov2008A,Mcauley2013Hidden}. However, none of them considers the pros and cons texts during the opinion modeling process, which is one of our major concern under the company profiling task.

Matrix factorization is a family of methods which is widely used for prediction. The intuition behind it is to get better data representation by projecting them into a latent space. Singular Value Decomposition (SVD)~\cite{golub1970singular} is a classic matrix factorization method for rating prediction, which gives low-rank approximations based on minimizing the sum-squared distance. However, since real-world data sets are often sparse, SVD does not perform well in practice. To solve it, some probabilistic matrix factorization methods have been proposed \cite{marlin2003modeling,marlin2004multiple,mnih2007probabilistic,Zeng-IJCAI-2015}. Probabilistic Matrix Factorization~\cite{mnih2007probabilistic} (PMF) is a representative one and has been popular in industry. However, in our salary prediction scenario, we need to model rating matrix and review text information simultaneously which cannot be met by neither SVD or PMF. Therefore, we develop a joint optimization framework to integrate the textual information (e.g., review) and numerical information (e.g., salary and rating) by extending Collaborative Topic Regression (CTR)~\cite{wang2011collaborative} for effective salary prediction.

\section{Preliminaries} \label{sec:propmodel}
In this section, we introduce some preliminaries used throughout this paper, including data description and problem definition.

\subsection{Data Description}

\begin{figure}[tb!]
\centering
\includegraphics[width=1.0\linewidth]{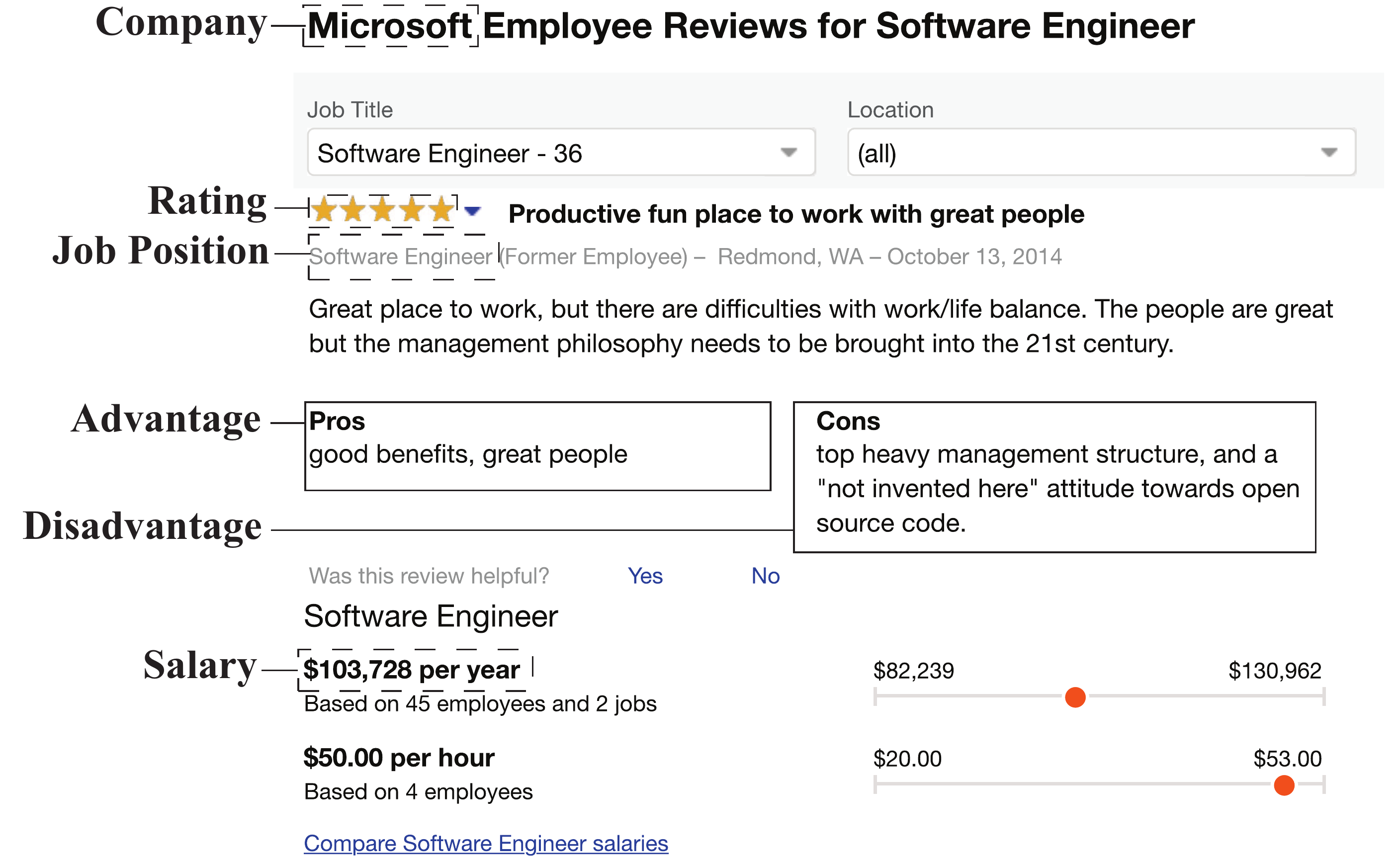}
\caption{An example of data description.}
\label{fig:indeed} 
\end{figure}

In this paper, we aim to leverage the data collected from online employment services for company profiling. To facilitate the understanding of our data, we show a page snapshot of Indeed\footnote{http://www.indeed.com/} in Figure~\ref{fig:indeed}. Specifically, each company has a number of reviews posted by its (former) employees, each of which contains the poster's job position (e.g., software engineer), textual information about the advantages and disadvantages of the company, and a rating score ranging from 1 to 5 to indicate the preferences of employees towards this company. Moreover, the salary range of each job position is also included for each company.

\subsection{Problem Statement}

Suppose we have a set of $E$ companies and a set of $J$ job positions. For each company $e$, there are many reviews referred to it. In each review, we have its reviewer's role (i.e., the reviewer's job position), rating, and two independent textual segments~(positive opinion and negative opinion). Moreover, we have the average salary for each job position. For simplicity, we group reviews by their job positions and denote two words lists as $\{w^{P}_{n,j,e}\}^{N}_{n=1}$ and $\{w^{C}_{m,j,e}\}^{M}_{m=1}$ to represent positive opinion and negative opinion for a specific job position~$j$, respectively.

Our problem is how to discover the latent representative patterns of job-company pair. To be more specific, there are two major tasks in this work: 1) how to learn positive and negative opinion patterns, ($\beta_j$, $\varphi_j$), for each job postition; and 2) how to use the latent patterns to predict job salaries ($\hat{s}_{j,e}$), for each job-company pair.

Thus, we propose a model, CPCTR, for jointly modeling the numerical information (i.e., rating and salary) and review content information simultaneously. To be more specific, we use probabilistic topic model to mine review content information and use matrix factorization to handle numerical information.
In terms of review content information, $\beta_j$ and $\varphi_j$ are represented by sets of opinion-related topics. Besides, each job-company pair ($j$, $e$) has a topic pattern $\theta_{j,e}$ indicating its probability over $\beta_j$ and $\varphi_j$.
In terms of numerical information, we use a low-dimensional representation derived from numerical information, such as salary and rating, to represent job position and combine it with $\theta_{j,e}$ to model the latent relationship among them.

Obviously, our model is a combination of probabilistic topic modeling and matrix factorization, similar to CTR. However, unlike CTR that only learns a global topic-word distribution $\beta$ and topic proportion $\theta_j$ for each item $j$, our model can learn two kinds of job related topic-word patterns, including a positive topic-word distribution $\beta_j$ and a negative topic-word distribution $\varphi_j$. Moreover, in contrast with CTR, which cannot incorporate both rating and salary information into one optimization model simultaneously, our method can model these two numerical values and utilize the learned opinion patterns for more precise salary prediction. Thus, our model leads to a more comprehensive interpretation of company profiling and provides a collaborative view from opinion modeling to salary prediction.

\begin{figure}[bt!]
\centering
\includegraphics[width=1\linewidth]{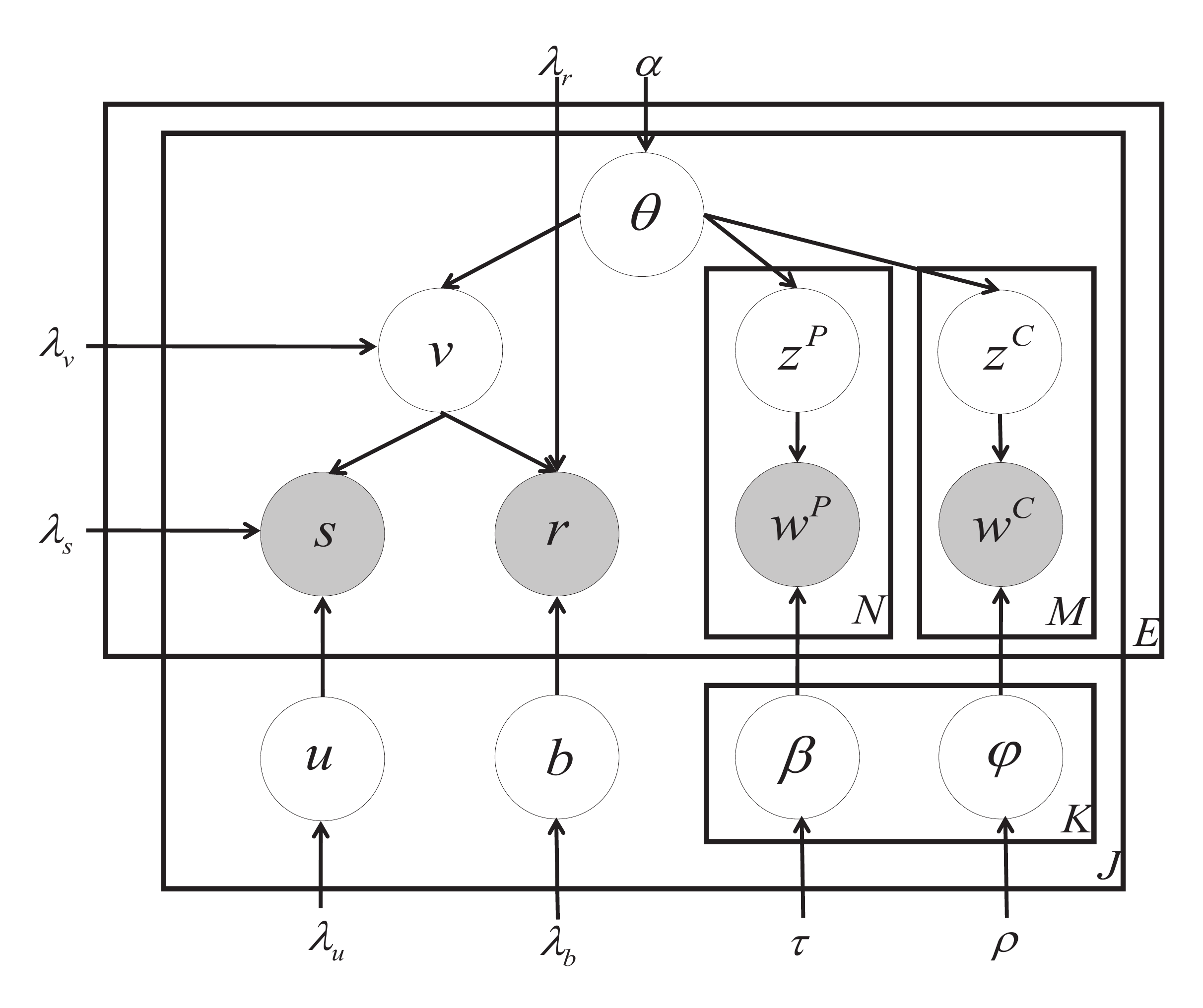}
\caption{The graphical representation of CPCTR.}
\label{fig:plate} 
\end{figure}

\section{Technical Details}
In this section, we formally introduce the technical details of our model CPCTR.

\subsection{Model Formulation}
As mentioned above, our model CPCTR is a Bayesian model which combines topic modeling with matrix factorization. The graphical representation of CPCTR is shown in Figure~\ref{fig:plate}. To facilitate understanding, we look into the model in two sides.

On one side, we model the job-company pair with a latent topic vector $\theta_{j,e} \in R^{K}$, where $K$ is the number of topics. In probabilistic topic modeling, job position $j$ can be represented by two latent matrices, i.e., the positive opinion topics $\beta_j \in R^{K \times G}$ and the negative opinion topics $\varphi_j \in R^{K \times G}$, where $G$ is the size of vocabulary.
For the n$th$ word~$w^{P}_{n,j,e}$ in a positive review of job-company pair, we assume there is a latent variable denoted as $z^{P}_{n,j,e} \in \{1,\ldots,K\}$, indicating the word's corresponding topic.
To be more specific, given $z^{P}_{n,j,e}=k$, $w^{P}_{n,j,e}$ follows a multinomial distribution parameterized by $\beta_{k,j}$.
Meanwhile, the positive latent pattern $z^{P}_{n,j,e}$ is considered to be drawn from the multinomial distribution $Mult(\theta_{j,e})$.
A similar process can be conducted for the negative review.

On the other side, we conduct matrix factorization for salary prediction.
In matrix factorization, we represent job position and job-company pair in a shared latent low-dimensional space of dimension $K$, i.e., job position $j$ is represented by latent vectors $u_{j} \in R^{K}$ and $b_{j} \in R^{K}$, which indicate the influences of job positions over salary and rating, respectively. Similarly, the job-company pair ($j, e$) is represented by a latent vector $v_{j,e} \in R^{K}$, which indicates the joint influences of job-company pair over numeric rating and salary values. Here, we assume the latent vector $u_j$ and $b_j$ follow Gaussian distributions with parameters $(0,\lambda^{-1}_{u}I_K)$ and $(0,\lambda^{-1}_{b}I_K)$, respectively. And, the latent vector $v_{j,e}$ is derived from $\theta_{j,e}$ by adding an offset, $\epsilon$. $\epsilon$ also follows a Gaussian distribution with parameters $(0,\lambda^{-1}_{v}I_K)$.
Therefore, it is obvious that $v_{j,e}$ is the key point by which we jointly model both content and numerical information.

We form the prediction of salary values of a specific job-company pair ($\hat{s}_{j,e}$) through the inner product between their latent representations, i.e.,
\begin{footnotesize}
\begin{align}\label{eq:salary}
&\hat{s}_{j,e}=u^{T}_{j}v_{j,e}.
\end{align}
\end{footnotesize}
Note that in our model, we first group reviews, ratings, and salary information by the posters' job-company pair. We then calculate the average ratings, average salaries and aggregate reviews as one single document for each job-company pair. The complete generative process of our model is demonstrated in Algorithm~\ref{algo:process}. In the following, we leverage the Bayesian approach for parameter learning.

\begin{footnotesize}
\renewcommand{\algorithmcfname}{ALGORITHM}
\begin{algorithm} [t]
\begin{enumerate}
\small
\item Draw $K*J$ topic patterns from its prior distribution,
  \begin{enumerate}
  \item Draw $\beta_{k,j}$ from the Dirichlet prior $Dir(\tau)$ for $k=1,\ldots,K$ and $j=1,\ldots,J$.
  \item Draw $\varphi_{k,j}$ from the Dirichlet prior $Dir(\rho)$ for $k=1,\ldots,K$ and $j=1,\ldots,J$.
  \end{enumerate}
\item For each job position $j$,
  \begin{enumerate}
  \item draw latent vector $u_j \sim N(0,\lambda^{-1}_{u}I_K)$.
  \item draw latent vector $b_j \sim N(0,\lambda^{-1}_{b}I_K)$.
  \end{enumerate}
\item For each job-company pair $(j, e)$,
  \begin{enumerate}
  \item Draw topic proportion $\theta_{j,e}$ from the Dirichlet prior $Dir(\alpha)$.
  \item For the n$th$ word $w^{P}_{n,j,e}$ of positive review,
    \begin{enumerate}
    \item Draw topic assignment $z^{P}_{n,j,e} \sim Mult(\theta_{j,e})$.
    \item Draw word $w^{P}_{n,j,e} \sim Mult(\beta_{z^{P}_{n,j,e},j})$.
    \end{enumerate}
  \item For the m$th$ word $w^{C}_{m,j,e}$ of negative review,
    \begin{enumerate}
    \item Draw topic assignment $z^{C}_{m,j,e} \sim Mult(\theta_{j,e})$.
    \item Draw word $w^{C}_{m,j,e} \sim Mult(\varphi_{z^{C}_{m,j,e},j})$.
    \end{enumerate}
  \item Draw latent offset $\epsilon_{j,e} \sim N(0,\lambda^{-1}_{v}I_K)$ and set latent vector $v_{j,e}=\epsilon_{j,e}+\theta_{j,e}$.
  \item Draw rating/salary values,
      \begin{enumerate}
      \item draw rating value $r_{j,e} \sim N(b^T_{j}v_{j,e},\lambda^{-1}_{r}I_K)$.
      \item draw salary value $s_{j,e} \sim N(u^T_{j}v_{j,e},\lambda^{-1}_{s}I_K)$.
      \end{enumerate}
  \end{enumerate}
\end{enumerate}
\caption{The Generative Process of CPCTR}\label{algo:process}
\end{algorithm}
\end{footnotesize}

\subsection{Parameter Learning}
In the above generative process, we denote mathmatical notations as follows. $S=\{s_{j,e}\}^{J,E}_{j=1,e=1}$, $R=\{r_{j,e}\}^{J,E}_{j=1,e=1}$, $W^P=\{w^{P}_{n,j,e}\}^{N,J,E}_{n=1,j=1,e=1}$, $W^C=\{w^{C}_{m,j,e}\}^{M,J,E}_{m=1,j=1,e=1}$, $\theta=\{\theta_{j,e}\}^{J,E}_{j=1,e=1}$, $U=\{u_{j}\}^{J}_{j=1}$, $B=\{b_{j}\}^{J}_{j=1}$, $V=\{v_{j,e}\}^{J,E}_{j=1,e=1}$. The joint likelihood of data, i.e., $S$, $R$, $W^P$, $W^C$, and the latent factors $\theta$, $\beta$, $\varphi$, $U$, $B$, $V$ under the full model is
\begin{footnotesize}
\begin{align}
& p(S, R, W^{P}, W^{C}, \theta, \beta, \varphi, U, B, V|\lambda_\bullet,\alpha,\tau,\rho)  \nonumber &\\
= & p(U|\lambda_u) p(B|\lambda_b) p(V|\lambda_v,\theta) p(R, S|U, B, V, \lambda_r, \lambda_s) \nonumber &\\
& p(W^P, W^C|\beta, \varphi, \theta) p(\theta|\alpha) p(\beta|\tau) p(\varphi|\rho) \nonumber &\\
= &\prod_j p(u_j|\lambda_u) \prod_j p(b_j|\lambda_b) \prod_e \prod_j p(v_{j,e}|\lambda_v,\theta_{j,e}) \nonumber &\\
& \prod_e \prod_j p(r_{j,e}|b_j, v_{j,e}, \lambda_r) \prod_e \prod_j p(s_{j,e}|u_j, v_{j,e}, \lambda_s) \nonumber &\\
& \prod_e \prod_j \left( \prod_n p(w^{P}_{n,j,e}|\beta_j,\theta_{j,e}) \prod_m p(w^{C}_{m,j,e}|\varphi_j,\theta_{j,e}) \right) \nonumber &\\
& p(\theta|\alpha) p(\beta|\tau) p(\varphi|\rho).
\end{align}
\end{footnotesize}
For learning the parameters, we develop an EM-style algorithm to learn the maximum a posterior (MAP) estimation. Maximization of posterior is equivalent to maximizing the complete log likelihood of $R$, $S$, $W^P$, $W^C$, $\theta$, $U$, $B$ and $V$, given $\lambda_\bullet$, $\beta$ and $\varphi$,

\begin{footnotesize}
\begin{align}
\mathcal{L}=&-\frac{\lambda_b}{2} \sum_j b_j^T b_j -\frac{\lambda_u}{2} \sum_j u_j^T u_j - \frac{\lambda_r}{2} \sum_e \sum_j (r_{j,e} - b^T_j v_{j,e})^2 \nonumber \\
&- \frac{\lambda_s}{2} \sum_e \sum_j (s_{j,e} - u^T_j v_{j,e})^2 - \frac{\lambda_v}{2}  \sum_e \sum_j (v_{j,e}-\theta_{j,e})^T  \nonumber \\
& (v_{j,e}-\theta_{j,e}) + \sum_e \sum_j \Bigg( \sum_n \log (\sum_k \theta_{k,j,e} \beta_{k,w^{P}_{n,j,e}}) \nonumber \\
& + \sum_m \log (\sum_k \theta_{k,j,e} \varphi_{k,w^{C}_{m,j,e}}) \Bigg).
\end{align}
\end{footnotesize}
Here, we employ coordinate ascent (CA) approach to alternatively optimize the latent factors \{$u_j$, $b_j$, $v_{j,e}$\} and the simplex variables $\theta_{j,e}$ as topic proportion. For $u_j$, $b_j$ and $v_{j,e}$, we follow in a similar fashion as for basic matrix factorization~\cite{hu2008collaborative}. Given the current estimation of $\theta_{j,e}$, taking the gradient of $\mathcal{L}$ with respect to $u_j$, $b_j$, $v_{j,e}$ and setting it to zero leads to
\begin{footnotesize}
\begin{align}
&u_j= (\lambda_u I_K+ \lambda_s \sum_e v_{j,e} v^T_{j,e})^{-1} \lambda_s \sum_e v_{j,e} s_{j,e} &\\
&b_j= (\lambda_b I_K+ \lambda_r \sum_e v_{j,e} v^T_{j,e})^{-1} \lambda_r \sum_e v_{j,e} r_{j,e} &\\
&v_{j,e}= (\lambda_v I_K + \lambda_s u_j u^T_j + \lambda_r b_j b^T_j)^{-1} &\nonumber\\
&\qquad(\lambda_v \theta_{j,e} + \lambda_s u_j s_{j,e} + \lambda_r b_j r_{j,e}). &
\end{align}
\end{footnotesize}
Given $U$, $B$ and $V$, we then apply a variational EM algorithm described in LDA~\cite{blei2003latent} to learn the topic proportion $\theta_{j,e}$.
We first define $q(z^{P}_{n,j,e}=k)=\phi^{P}_{k,n,j,e}$ and $q(z^{C}_{m,j,e}=k)=\phi^{C}_{k,m,j,e}$, and then we separate the items that contain $\theta_{j,e}$ and apply Jensen's inequality,
\begin{footnotesize}
\begin{align}\label{eq:Ltheta}
\mathcal{L}(\theta_{j,e}) >=& - \frac{\lambda_v}{2} (v_{j,e}-\theta_{j,e})^T (v_{j,e}-\theta_{j,e}) + \nonumber &\\
&\sum_n \sum_k \phi^{P}_{k,n,j,e} (\log \theta_{k,j,e} \beta_{k,w^{P}_{n,j,e}} - \log \phi^{P}_{k,n,j,e}) + \nonumber &\\
&\sum_m \sum_k \phi^{C}_{k,m,j,e} (\log \theta_{k,j,e} \varphi_{k,w^{C}_{m,j,e}} - \log \phi^{C}_{k,m,j,e}) \nonumber &\\
=& \mathcal{L}(\theta_{j,e},\phi_{j,e}), &
\end{align}
\end{footnotesize}
where $\phi_{j,e}=\left\{\{\phi^{P}_{k,n,j,e}\}^{K,N}_{k=1,n=1}, \{\phi^{C}_{k,m,j,e}\}^{K,M}_{k=1,m=1} \right\}$. In the E-step, the optimal variational multinomial $\phi^{P}_{k,n,j,e}$ and $\phi^{C}_{k,m,j,e}$ satisfy
\begin{footnotesize}
\begin{align}
&\phi^{P}_{k,n,j,e} \propto \theta_{k,j,e} \beta_{k,w^{P}_{n,j,e}} \\
&\phi^{C}_{k,m,j,e} \propto \theta_{k,j,e} \varphi_{k,w^{C}_{m,j,e}}.
\end{align}
\end{footnotesize}
The $\mathcal{L}(\theta_{j,e},\phi_{j,e})$ gives a tight lower bound of $\mathcal{L}(\theta_{j,e})$. Similar to CTR~\cite{wang2011collaborative}, we use projection gradient \cite{bertsekas1999nonlinear} to optimize $\theta_{j,e}$. Coordinate ascent can be applied to optimize remaining parameters $U$, $B$, $V$, $\theta$ and $\phi$. Then following the same M-step for topics in LDA~\cite{blei2003latent}, we optimize $\beta$ and $\varphi$ as follows,
\begin{footnotesize}
\begin{align}
& \beta_{g,k,j} \propto \sum_e \sum_n \phi^{P}_{k,n,j,e} 1[w^{P}_{n,j,e}=g] \\
& \varphi_{g,k,j} \propto \sum_e \sum_m \phi^{C}_{k,m,j,e} 1[w^{C}_{m,j,e}=g],
\end{align}
\end{footnotesize}
where we denote $g$ as an arbitrary term in the vocabulary set.

\subsection{Discussion on Salary Prediction}
After all the  optimal parameters are learned, the CPCTR model can be used for salary prediction by Equation~\ref{eq:salary}. In this task, rating values and review content of the predicted job-company pair ($j$, $e$) are available, but no salary information of ($j$, $e$) pair is available. To obtain the topic proportion $\theta^*_{j,e}$ for the predicted job-company pair ($j$, $e$), we optimize Equation~\ref{eq:Ltheta}.

In particular, we only focus on the task of salary prediction, although rating prediction can be conducted in a similar way. Since reviews are always accompanied by ratings, ratings should be regarded as part of opinion information. Therefore, in this work we treat the ratings as the complementary of reviews for opinion mining, and the side information for salary prediction.



\section{Experimental Results}\label{sec:results}
In this section, we first give a short parameter sensitivity discussion to show the robustness of our model and then evaluate the salary prediction performance of CPCTR based on a real-world data set with several state-of-the-art baselines. Finally, we empirically study the pros and cons for each job-company pair learned from their employees' review.

\subsection{Experimental Setup}
\paragraph{Data Sets.}
Kanzhun\footnote{http://www.kanzhun.com/} is one the largest online employment website in China, where members can review companies and assign numeric ratings from 1 to 5, and post their own salary information. Thus, Kanzhun provides an ideal data source for experiments on company profiling and salary prediction. The data set used in our experiments consists of 934 unique companies which contains at least one of total 1,128 unique job positions, i.e., for a specific company, at least one job's average salary and rating has been included. Moreover, the data set contains 4,682 average salaries for all job-company pair (the matrix has a sparsity of 99.6\%). The average rating and average salary in our data set are 3.32 and 7,565.21, respectively. For each review, we extracted advantages and disadvantages, then grouped reviews by its job position and formed one document for each job-company pair. Particularly, we removed stop words and single words, filtered out words that appear in less than one document and more than 90\% of documents and then choose only the first 10,000 most frequent words as the vocabulary, which yielded a corpus of 580K negative words and 652K positive words. Finally, we converted documents into the bag-of-words format for model learning.

\paragraph{Baseline Methods.}
To evaluate the performance of salary prediction for CPCTR, we chose three state-of-the-art benchmark methods for comparisons, including \textbf{PMF}~\cite{mnih2007probabilistic}, Regularized Singular Value Decomposition of data with missing values \textbf{RSVD}\footnote{https://github.com/alabid/PySVD} and Collaborative Topic Regression \textbf{CTR}~\cite{wang2011collaborative}.

\paragraph{Evaluation Metrics.}
We used two widely-used metrics, i.e., root Mean Square Error (rMSE), Mean Absolute Error (MAE), for measuring the prediction performance of different models. Specifically, we have
\begin{footnotesize}
\begin{align}
& rMSE = \sqrt[]{\frac{1}{N} \sum_i (s_i - \hat{s_i})^2} & \\
& MAE = \frac{1}{N} \sum_i |s_i - \hat{s_i}|, &
\end{align}
\end{footnotesize}
where $s_i$ is the actual salary of $i$th job-company pair, $\hat{s_i}$ is its salary prediction and $N$ is the number of test instances.

\begin{table}[bt!]
  \centering
  \footnotesize
  \caption{The prediction performance of different methods under 5-fold cross-validation.}
    \resizebox{0.8\linewidth}{!}{
    \begin{tabular}{ccrrrr}
    \toprule
    Fold  & \multicolumn{1}{l}{Method} & CPCTR & PMF   & RSVD  & CTR \\
    \midrule
    \multirow{2}[2]{*}{1} & \multicolumn{1}{l}{rMSE} & \textbf{0.0528} & \textit{0.0561} & 0.0608 & 0.0670 \\
          & \multicolumn{1}{l}{MAE} & \textbf{0.0347} & \textit{0.0356} & 0.0419 & 0.0433 \\
    \midrule
    \multirow{2}[2]{*}{2} & \multicolumn{1}{l}{rMSE} & \textit{0.0530} & \textbf{0.0518} & 0.0592 & 0.0597 \\
          & \multicolumn{1}{l}{MAE} & \textit{0.0346} & \textbf{0.0332} & 0.0401 & 0.0394 \\
    \midrule
    \multirow{2}[2]{*}{3} & \multicolumn{1}{l}{rMSE} & \textit{0.0506} & \textbf{0.0499} & 0.0595 & 0.0621 \\
          & \multicolumn{1}{l}{MAE} & \textit{0.0328} & \textbf{0.0322} & 0.0413 & 0.0414 \\
    \midrule
    \multirow{2}[2]{*}{4} & \multicolumn{1}{l}{rMSE} & \textbf{0.0680} & \textit{0.0703} & 0.0743 & 0.0815 \\
          & \multicolumn{1}{l}{MAE} & \textbf{0.0345} & \textit{0.0365} & 0.0425 & 0.0472 \\
    \midrule
    \multirow{2}[2]{*}{5} & \multicolumn{1}{l}{rMSE} & \textbf{0.0479} & \textit{0.0514} & 0.0543 & 0.0609 \\
          & \multicolumn{1}{l}{MAE} & \textbf{0.0332} & \textit{0.0354} & 0.0407 & 0.0433 \\
    \midrule
    \multicolumn{2}{r}{Average rMSE} & \textbf{0.0545} & \textit{0.0559} & 0.0616 & 0.0662 \\
    \multicolumn{2}{r}{Average MAE} & \textbf{0.0340} & \textit{0.0346} & 0.0413 & 0.0429 \\
    \bottomrule
    \end{tabular}}
  \label{tab:pred_perfo}%
\end{table}%

\paragraph{Experimental Settings.}
In our experiments, we used 5-fold cross-validation. For every job position that was posted by at least 5 companies, we evenly split their job-company pairs (average rating/salary values) into 5 folds. We iteratively considered each fold to be a test set and the others to be the training set. For those job positions that were posted by fewer than 5 companies, we always put them into the training set. This leads to that all job positions in the test set must have appeared in the training set, thus it guarantees the in-matrix scenario for CTR model in prediction. For each fold, we fitted a model to the training set and test on the within-fold jobs for each company. Note that, each company has a different set of within-fold jobs. Finally, we obtained the predicted salaries and evaluated them on the test set.

The parameter settings of different methods are stated as follows. For all methods, we set the number of latent factor to $K=5$ and the maximum iterations for convergence as $max\_iter=500$. For probabilistic topic modeling in CTR and CPCTR, we set the parameters $\alpha_{smooth}=0.01$. For CTR, we used fivefold cross validation to find that $\lambda_u=10$, $\lambda_v=0.01$, $a=1$ and $b=0.01$ provides the best performance. For our model CPCTR, we chose the parameters by using grid search on held out predictions. As a default setting for CPCTR, we set $\lambda_u=0.1$, $\lambda_b=0.01$, $\lambda_v=1000$, $\lambda_r=1$, $\lambda_s=1$. More detailed discussions about parameter sensitivity of our model will be given in the following subsection. Additionally, for convenience of parameter choosing, we used min-max method to normalize all values of rating/salary into [0, 1] range.

\begin{figure}[bt!]
\centering
    \begin{subfigure}[b]{0.5\linewidth}
      \includegraphics[width=1.0\linewidth]{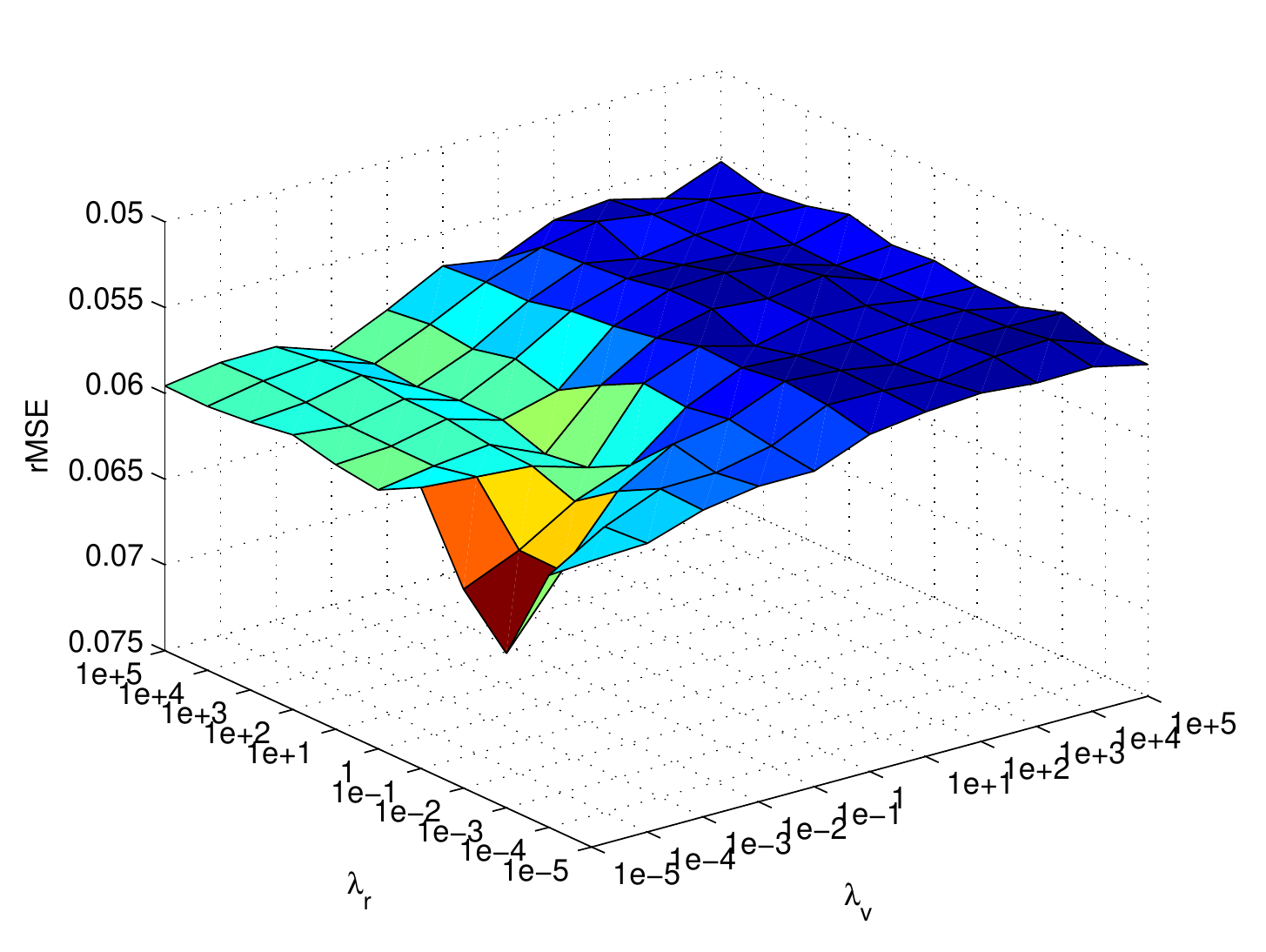}
    \end{subfigure}%
    \begin{subfigure}[b]{0.5\linewidth}
      \includegraphics[width=1.0\linewidth]{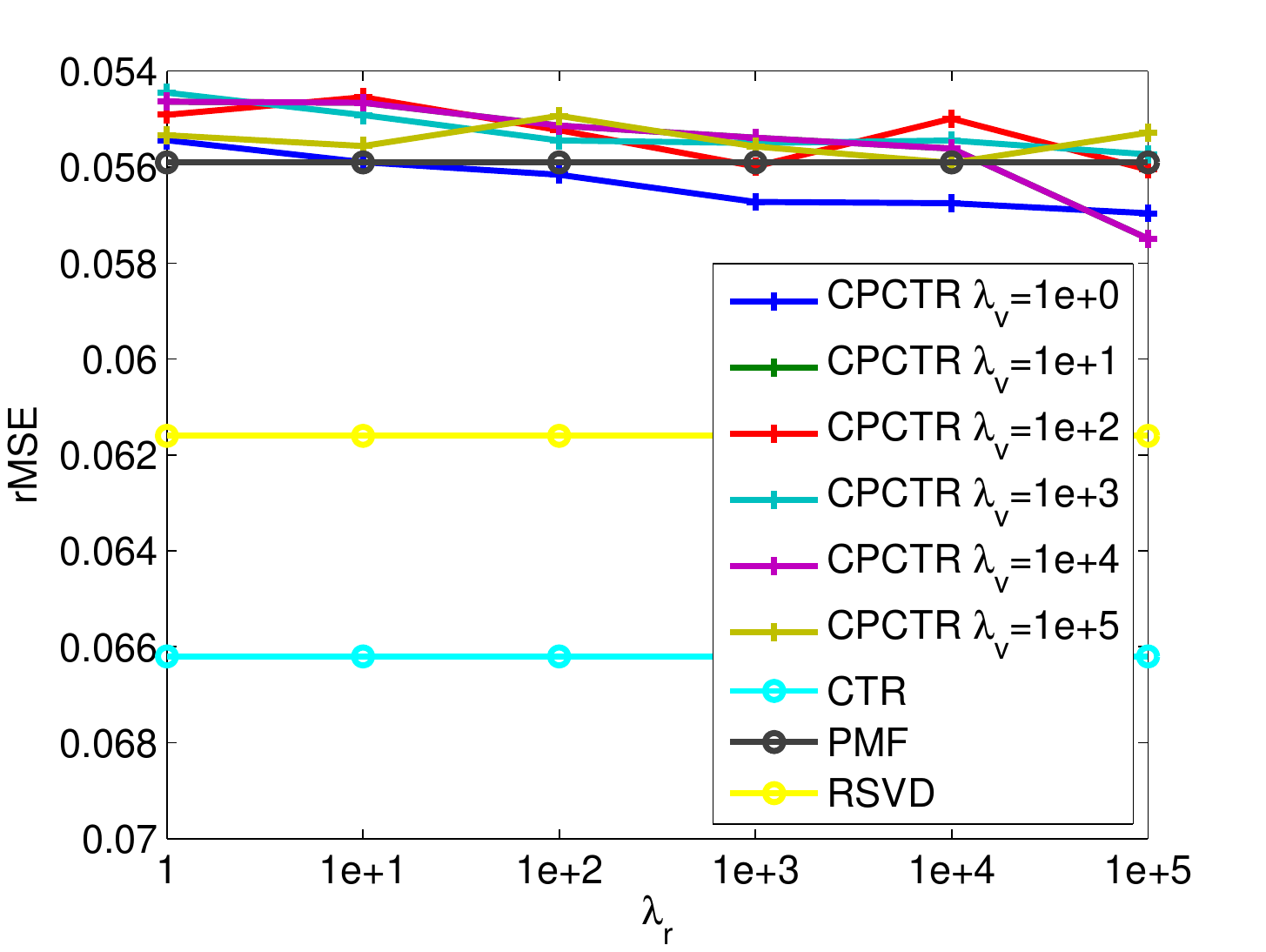}
    \end{subfigure}%
  \caption{Plots of prediction performance for CPCTR by varying content parameter $\lambda_v$ and rating parameter $\lambda_r$.}
  \label{fig:parasens}
\end{figure}

\begin{figure}[bt!]
	\centering
	\includegraphics[width=0.8\linewidth]{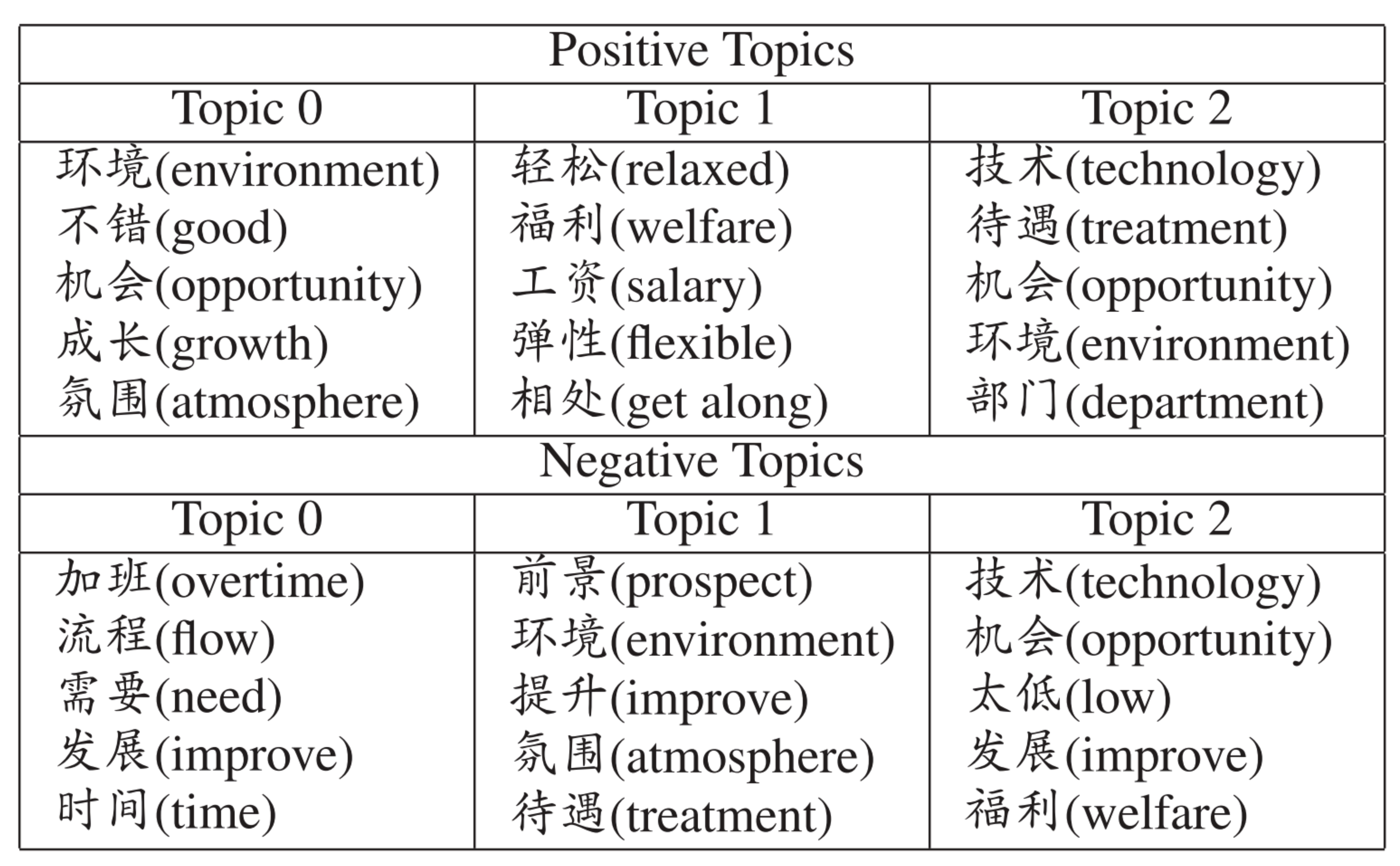}
	\caption{Topics of job position ``software engineer''.}
	\label{fig:job_topics} 
\end{figure}

\begin{figure*}[t!]
	\centering
	\includegraphics[width=14cm]{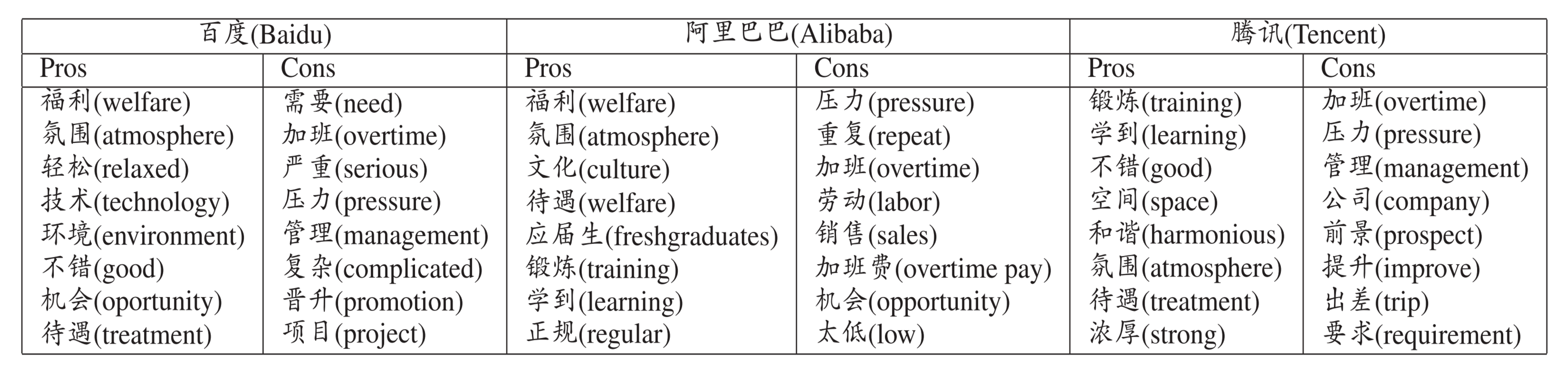}
	\caption{Pros and cons of various enterprises given job position ``software engineer''.}
	\label{fig:e_pros_cons} 
\end{figure*}

\subsection{Parameter Sensitivity}
In our model, the content parameter $\lambda_v$ controls the contribution of review content information to model performance and the rating parameter $\lambda_r$ balances the contribution of rating information to model performance. In the left plot of Figure~\ref{fig:parasens}, we vary the content parameter $\lambda_v$ and rating parameter $\lambda_r$ from $1e-5$ to $1e+5$ to study the effect on the performance of salary prediction, and the average performance within fivefold cross validation is shown in this plot. First, we can see that CPCTR shows good prediction performance when $\lambda_v>=1$ and $\lambda_r>=1$, and achieves the best prediction accuracy when $\lambda_v=1000$ and $\lambda_r=1$, which is the default setting for CPCTR. Next, for facilitating comparison, we shrink the range of $\lambda_v$ and $\lambda_r$ into [1, 1e+05] and show the right plot of Figure~\ref{fig:parasens}. From this plot, we can see that almost all cases of CPCTR outperform other state-of-the-art baselines, except for $\lambda_v=1$. The results show small and negligible fluctuation with varied $\lambda_v$ and $\lambda_r$, and CPCTR becomes insensitive to these two parameters.

\subsection{Performance of Salary Prediction}

We show the prediction performance of different methods in Table~\ref{tab:pred_perfo}. Note that, the best results are highlighted in bold and the runner-up are denoted in italic. From the results, we could observe that CPCTR achieves the best average prediction performance in terms of 5-fold cross-validation, and outperforms other baselines in three folds. This is in great contrast to CTR, which shows poor prediction performance in all five folds. It is because that, although CTR can integrate textual information for salary prediction, it cannot utilize the rating information and does not explicitly model the positive/negtive topic-word distribution. Among traditional collaborative filtering methods, PMF consistently outperforms RSVD in all five folds, which demonstrates the effectiveness of probabilistic methods on prediction tasks. Based on the above analysis, CPCTR can be regarded as a more comprehensive and effective framework for company profiling, that can integrate review opinions and ratings for salary prediction.

\subsection{Empirical Study of Opinion Profiling}
Here, we apply CPCTR to carry out opinion analysis for different companies based on employees' reviews. The objective is to effectively reveal the pros and cons of companies, which indeed helps for competitor benchmarking.

To illustrate the effectiveness of learning job-position level topic-word distributions, we listed 3 positive topics and 3 negative topics of job position \textit{Software Engineer} inferred from CPCTR, as shown in Figure~\ref{fig:job_topics}. Each topic is represented by 5 most probable words for that topic. It can be seen that our method has an effective interpretation of latent job position pattern and these topics accurately capture the common semantics of the job position \textit{Software Engineer} in the whole market. We can see some interesting postive/negative topic patterns. For positive topics, topic 0 is about job environment, topic 1 is about flexible work time, topic 2 is technology atmosphere. For negative topics, topic 0 is about overtime, topic 1 is about prospect and promotion, and topic 2 is about opportunity and welfare.

We also compared the pros and cons among BAT, which is the abbreviation of three largest and most representative Chinese Internet companies, i.e., Baidu, Alibaba and Tencent. Specifically, we presented the pros and cons with most probable words appearing in learned topics for each company, given the job position \textit{Software Engineer} in Figure~\ref{fig:e_pros_cons}. As can be seen, topics for each job-company pair can effectively capture the specific characteristics of each company. For instance, the typical pros of Baidu are welfare and technology, while those of Tencent are training and learning and those of Alibaba are culture and atmosphere. Interestingly, employees of all these three companies chose overtime as their cons, and the management of Tencent seems to be a typical cons.

\section{Concluding Remarks}\label{sec:conclusion}
In this paper, we proposed a model CPCTR for company profiling, which can collaboratively model the textual information and numerical information of companies. A unique perspective of CPCTR is that it formulates a joint optimization framework for learning the latent patterns of companies, including the positive/negative opinions of companies and the latent topic variable that influences salary from an employee's perspective. With the identified patterns, both opinion analysis and salary prediction can be conducted effectively. Finally, we conducted extensive experiments on a real-world data set. The results showed that our model provides a comprehensive interpretation of company characteristics and a more effective salary prediction than baselines.

\section{Acknowledgments}
This work was partially supported by NSFC (71322104, 71531001, 71471009, 71490723), National High Technology Research and Development Program of China (SS2014AA012303), National Center for International Joint Research on E-Business Information Processing (2013B01035), and Fundamental Research Funds for the Central Universities.

\bibliography{aaaiproc}

\begin{thebibliography}{}

\bibitem[\protect\citeauthoryear{Alling}{2002}]{alling2002method}
Alling, E.
\newblock 2002.
\newblock Method and system for facilitating multi-enterprise benchmarking
  activities and performance analysis.
\newblock US Patent App. 10/137,218.

\bibitem[\protect\citeauthoryear{Bertsekas}{1999}]{bertsekas1999nonlinear}
Bertsekas, D.
\newblock 1999.
\newblock {\em Nonlinear Programming}.
\newblock Athena Scientific.

\bibitem[\protect\citeauthoryear{Blei and
  Lafferty}{2005}]{lafferty2005correlated}
Blei, D.~M., and Lafferty, J.~D.
\newblock 2005.
\newblock Correlated topic models.
\newblock In {\em Proceedings of the 18th International Conference on Neural
  Information Processing Systems}, NIPS'05,  147--154.
\newblock Cambridge, MA, USA: MIT Press.

\bibitem[\protect\citeauthoryear{Blei and Lafferty}{2006}]{blei2006dynamic}
Blei, D.~M., and Lafferty, J.~D.
\newblock 2006.
\newblock Dynamic topic models.
\newblock In {\em Proceedings of the 23rd International Conference on Machine
  Learning}, ICML '06,  113--120.
\newblock New York, NY, USA: ACM.

\bibitem[\protect\citeauthoryear{Blei, Ng, and Jordan}{2003}]{blei2003latent}
Blei, D.~M.; Ng, A.~Y.; and Jordan, M.~I.
\newblock 2003.
\newblock Latent dirichlet allocation.
\newblock {\em J. Mach. Learn. Res.} 3:993--1022.

\bibitem[\protect\citeauthoryear{Ganu, Elhadad, and
  Marian}{2009}]{Ganu2009Beyond}
Ganu, G.; Elhadad, N.; and Marian, A.
\newblock 2009.
\newblock Beyond the stars: Improving rating predictions using review text
  content.
\newblock In {\em 12th International Workshop on the Web and Databases, WebDB
  2009, Providence, Rhode Island, USA, June 28, 2009}.

\bibitem[\protect\citeauthoryear{Golub and Reinsch}{1970}]{golub1970singular}
Golub, G.~H., and Reinsch, C.
\newblock 1970.
\newblock Singular value decomposition and least squares solutions.
\newblock {\em Numer. Math.} 14(5):403--420.

\bibitem[\protect\citeauthoryear{Hofmann}{1999}]{hofmann1999probabilistic}
Hofmann, T.
\newblock 1999.
\newblock Probabilistic latent semantic indexing.
\newblock In {\em Proceedings of the 22Nd Annual International ACM SIGIR
  Conference on Research and Development in Information Retrieval}, SIGIR '99,
  50--57.
\newblock New York, NY, USA: ACM.

\bibitem[\protect\citeauthoryear{Hollocks \bgroup et al\mbox.\egroup
  }{1997}]{hollocks1997assessing}
Hollocks, B.~W.; Goranson, H.~T.; Shorter, D.~N.; and Vernadat, F.~B.
\newblock 1997.
\newblock {\em Assessing Enterprise Integration for Competitive
  Advantage---Workshop 2, Working Group 1}.
\newblock Berlin, Heidelberg: Springer Berlin Heidelberg.
\newblock  96--107.

\bibitem[\protect\citeauthoryear{Hu, Koren, and
  Volinsky}{2008}]{hu2008collaborative}
Hu, Y.; Koren, Y.; and Volinsky, C.
\newblock 2008.
\newblock Collaborative filtering for implicit feedback datasets.
\newblock In {\em 2008 Eighth IEEE International Conference on Data Mining},
  263--272.

\bibitem[\protect\citeauthoryear{Kerschbaum}{2008}]{kerschbaum2008building}
Kerschbaum, F.
\newblock 2008.
\newblock Building a privacy-preserving benchmarking enterprise system.
\newblock {\em Enterprise Information Systems} 2(4):421--441.

\bibitem[\protect\citeauthoryear{Knuf}{2000}]{knuf2000benchmarking}
Knuf, J.
\newblock 2000.
\newblock Benchmarking the lean enterprise: Organizational learning at work.
\newblock {\em Journal of Management in Engineering} 16(4):58--71.

\bibitem[\protect\citeauthoryear{Marlin and Zemel}{2004}]{marlin2004multiple}
Marlin, B., and Zemel, R.~S.
\newblock 2004.
\newblock The multiple multiplicative factor model for collaborative filtering.
\newblock In {\em Proceedings of the Twenty-first International Conference on
  Machine Learning}, ICML '04,  73--.
\newblock New York, NY, USA: ACM.

\bibitem[\protect\citeauthoryear{Marlin}{2003}]{marlin2003modeling}
Marlin, B.
\newblock 2003.
\newblock Modeling user rating profiles for collaborative filtering.
\newblock In {\em Proceedings of the 16th International Conference on Neural
  Information Processing Systems}, NIPS'03,  627--634.
\newblock Cambridge, MA, USA: MIT Press.

\bibitem[\protect\citeauthoryear{Martin and Rice}{2007}]{martin2007profiling}
Martin, N.~J., and Rice, J.~L.
\newblock 2007.
\newblock Profiling enterprise risks in large computer companies using the
  leximancer software tool.
\newblock {\em Risk Management} 9(3):188--206.

\bibitem[\protect\citeauthoryear{McAuley and
  Leskovec}{2013}]{Mcauley2013Hidden}
McAuley, J., and Leskovec, J.
\newblock 2013.
\newblock Hidden factors and hidden topics: Understanding rating dimensions
  with review text.
\newblock In {\em Proceedings of the 7th ACM Conference on Recommender
  Systems}, RecSys '13,  165--172.
\newblock New York, NY, USA: ACM.

\bibitem[\protect\citeauthoryear{Rosen-Zvi \bgroup et al\mbox.\egroup
  }{2004}]{rosen2004author}
Rosen-Zvi, M.; Griffiths, T.; Steyvers, M.; and Smyth, P.
\newblock 2004.
\newblock The author-topic model for authors and documents.
\newblock In {\em Proceedings of the 20th Conference on Uncertainty in
  Artificial Intelligence}, UAI '04,  487--494.
\newblock Arlington, Virginia, United States: AUAI Press.

\bibitem[\protect\citeauthoryear{Salakhutdinov and
  Mnih}{2007}]{mnih2007probabilistic}
Salakhutdinov, R., and Mnih, A.
\newblock 2007.
\newblock Probabilistic matrix factorization.
\newblock In {\em Proceedings of the 20th International Conference on Neural
  Information Processing Systems}, NIPS'07,  1257--1264.
\newblock USA: Curran Associates Inc.

\bibitem[\protect\citeauthoryear{Seong~Leem \bgroup et al\mbox.\egroup
  }{2008}]{seong2008information}
Seong~Leem, C.; Wan~Kim, B.; Jung~Yu, E.; and Ho~Paek, M.
\newblock 2008.
\newblock Information technology maturity stages and enterprise benchmarking:
  an empirical study.
\newblock {\em Industrial Management \& Data Systems} 108(9):1200--1218.

\bibitem[\protect\citeauthoryear{Titov and McDonald}{2008}]{Titov2008A}
Titov, I., and McDonald, R.
\newblock 2008.
\newblock A joint model of text and aspect ratings for sentiment summarization.
\newblock In {\em Proceedings of ACL-08: HLT},  308--316.
\newblock Columbus, Ohio: Association for Computational Linguistics.

\bibitem[\protect\citeauthoryear{Vivekanandan and
  Aravindan}{2014}]{K2014Aspect}
Vivekanandan, K., and Aravindan, J.~S.
\newblock 2014.
\newblock Aspect-based opinion mining: A survey.
\newblock {\em International Journal of Computer Applications} 106(3):21--26.

\bibitem[\protect\citeauthoryear{Wang and Blei}{2011}]{wang2011collaborative}
Wang, C., and Blei, D.~M.
\newblock 2011.
\newblock Collaborative topic modeling for recommending scientific articles.
\newblock In {\em Proceedings of the 17th ACM SIGKDD International Conference
  on Knowledge Discovery and Data Mining}, KDD '11,  448--456.
\newblock New York, NY, USA: ACM.

\bibitem[\protect\citeauthoryear{Zeng \bgroup et al\mbox.\egroup
  }{2015}]{Zeng-IJCAI-2015}
Zeng, G.; Zhu, H.; Liu, Q.; Luo, P.; Chen, E.; and Zhang, T.
\newblock 2015.
\newblock Matrix factorization with scale-invariant parameters.
\newblock In {\em Proceedings of the Twenty-Fourth International Joint
  Conference on Artificial Intelligence, {IJCAI} 2015, Buenos Aires, Argentina,
  July 25-31, 2015},  4017--4024.

\bibitem[\protect\citeauthoryear{Zhu \bgroup et al\mbox.\egroup
  }{2014}]{Zhu-ICDM-2014}
Zhu, C.; Zhu, H.; Ge, Y.; Chen, E.; and Liu, Q.
\newblock 2014.
\newblock Tracking the evolution of social emotions: {A} time-aware topic
  modeling perspective.
\newblock In {\em 2014 {IEEE} International Conference on Data Mining, {ICDM}
  2014, Shenzhen, China, December 14-17, 2014},  697--706.

\bibitem[\protect\citeauthoryear{Zhu \bgroup et al\mbox.\egroup
  }{2016}]{Zhu-KDD-2016}
Zhu, C.; Zhu, H.; Xiong, H.; Ding, P.; and Xie, F.
\newblock 2016.
\newblock Recruitment market trend analysis with sequential latent variable
  models.
\newblock In {\em Proceedings of the 22nd {ACM} {SIGKDD} International
  Conference on Knowledge Discovery and Data Mining, {KDD} 2016, San Francisco,
  CA, USA, August 13-17, 2016},  383--392.

\end{thebibliography}
\bibliographystyle{aaai}

\end{document}